\documentclass[aps,prl,imp,twocolumn,superscriptaddress]{revtex4}
\usepackage[pdftex,
 letterpaper=true,
 hyperindex=true,
 breaklinks=true,
 colorlinks=false,
 citecolor=blue,
 pdftitle={},
 pdfauthor={}]
{hyperref}
\usepackage{graphicx}
\begin{document}

\title{Evidence for the ground-state resonance of $^{26}$O}

\author{E.~Lunderberg}
	\affiliation{Department of Physics, Hope College, Holland, Michigan 49423, USA}
\author{P.A.~DeYoung}
	\affiliation{Department of Physics, Hope College, Holland, Michigan 49423, USA}
\author{Z.~Kohley}
	\affiliation{National Superconducting Cyclotron Laboratory, Michigan State University, East Lansing, Michigan 48824, USA}
\author{H.~Attanayake}
    \affiliation{Department of Physics \& Astronomy, Ohio University, Athens, Ohio 45701, USA}
\author{T.~Baumann}
	\affiliation{National Superconducting Cyclotron Laboratory, Michigan State University, East Lansing, Michigan 48824, USA}
\author{D.~Bazin}
	\affiliation{National Superconducting Cyclotron Laboratory, Michigan State University, East Lansing, Michigan 48824, USA}
\author{G.~Christian}
	\affiliation{National Superconducting Cyclotron Laboratory, Michigan State University, East Lansing, Michigan 48824, USA}
	\affiliation{Department of Physics \& Astronomy, Michigan State University, East Lansing, Michigan 48824, USA}
\author{D.~Divaratne}
    \affiliation{Department of Physics \& Astronomy, Ohio University, Athens, Ohio 45701, USA}
\author{S.M.~Grimes}
    \affiliation{Department of Physics \& Astronomy, Ohio University, Athens, Ohio 45701, USA}
\author{A.~Haagsma}
    \affiliation{Department of Physics, Central Michigan University, Mt. Pleasant, Michigan, 48859, USA}
\author{J.E.~Finck}
    \affiliation{Department of Physics, Central Michigan University, Mt. Pleasant, Michigan, 48859, USA}
\author{N.~Frank}
	\affiliation{Department of Physics \& Astronomy, Augustana College, Rock Island, Illinois, 61201, USA}
\author{B.~Luther}
    \affiliation{Department of Physics, Concordia College, Moorhead, Minnesota 56562, USA}
\author{S.~Mosby}
	\affiliation{National Superconducting Cyclotron Laboratory, Michigan State University, East Lansing, Michigan 48824, USA}
	\affiliation{Department of Physics \& Astronomy, Michigan State University, East Lansing, Michigan 48824, USA}
\author{T.~Nagy}
	\affiliation{Department of Physics, Hope College, Holland, Michigan 49423, USA}
\author{G.F.~Peaslee}
	\affiliation{Department of Physics, Hope College, Holland, Michigan 49423, USA}
\author{A.~Schiller}
    \affiliation{Department of Physics \& Astronomy, Ohio University, Athens, Ohio 45701, USA}
\author{J.~Snyder}
	\affiliation{National Superconducting Cyclotron Laboratory, Michigan State University, East Lansing, Michigan 48824, USA}
	\affiliation{Department of Physics \& Astronomy, Michigan State University, East Lansing, Michigan 48824, USA}
\author{A.~Spyrou}
	\affiliation{National Superconducting Cyclotron Laboratory, Michigan State University, East Lansing, Michigan 48824, USA}
	\affiliation{Department of Physics \& Astronomy, Michigan State University, East Lansing, Michigan 48824, USA}
\author{M.J.~Strongman}
	\affiliation{National Superconducting Cyclotron Laboratory, Michigan State University, East Lansing, Michigan 48824, USA}
	\affiliation{Department of Physics \& Astronomy, Michigan State University, East Lansing, Michigan 48824, USA}
\author{M.~Thoennessen}
    \email[]{thoennessen@nscl.msu.edu}
	\affiliation{National Superconducting Cyclotron Laboratory, Michigan State University, East Lansing, Michigan 48824, USA}
	\affiliation{Department of Physics \& Astronomy, Michigan State University, East Lansing, Michigan 48824, USA}

\date{\today}

\begin{abstract}
Evidence for the ground state of the neutron-unbound nucleus $^{26}$O was observed for the first time in the single proton-knockout reaction from a 82~MeV/u $^{27}$F beam. Neutrons were measured in coincidence with $^{24}$O fragments. $^{26}$O was determined to be unbound by 150$^{+50}_{-150}$~keV from the observation of low-energy neutrons. This result agrees with recent shell model calculations based on microscopic two- and three-nucleon forces.


\end{abstract}

\maketitle

A major challenge in nuclear physics remains the description of nuclei based on fundamental interactions. ``Ab-initio'' approaches have been developed to calculate nuclear properties based on nucleon--nucleon scattering data up to A $\sim$ 12 \cite{2001Pie01}. Recent advances in nuclear theory made it possible to describe some fundamental properties of light nuclei up to oxygen based on two- and three-nucleon interactions \cite{2006Gou01,2007Nav01,2010Ots01,2010Cor01,2011Jen01}. On the way to heavier nuclides it will be critical for these models to describe the dramatic change in the location of the neutron dripline from oxygen ($N = 16$) to fluorine ($N \ge 22$) which was first pointed out by Sakurai et al.  \cite{1999Sak01}. The addition of one proton binds at least six additional neutrons. The two-neutron separation energy of $^{26}$O serves as an important benchmark for these calculations. The majority of the current nuclear structure models predict $^{26}$O to be bound \cite{2008Nak01,2008Sch02,2008Sch01,2009Sat01,2009Mas01,2011Shu01}. Experimentally it has been shown that $^{24}$O is bound \cite{1970Art01} while repeated searches for bound $^{25}$O and $^{26}$O have been unsuccessful \cite{2012Tho01,1985Lan01,1997Tar01,2003Tho01,1990Gui01,1996Fau01,2005Sch01}, although $^{25}$O had initially been reported as being particle stable \cite{1981Ste01}. Shell-model calculations using phenomenological interactions do predict $^{26}$O to be unbound: SDPF-M \cite{1999Uts01} by 77~keV and USD05a \cite{2006Bro01} by 510~keV. A continuum shell model calculation predicts $^{26}$O to be unbound by only 21~keV \cite{2006Vol01}. Recently it was shown that three-body forces are necessary to describe the binding energies of neutron-rich oxygen isotopes based on fundamental nucleon--nucleon forces \cite{2010Ots01}. However, no calculations have been published which simultaneously predict $^{26}$O to be unbound and $^{31}$F to be bound.

Because no bound states of $^{26}$O exist, the search for its elusive ground state must be extended to unbound states. The unbound ground state of $^{25}$O was measured using invariant mass spectroscopy and was found to have a decay energy of 770$^{+20}_{-10}$~keV \cite{2008Hof01}. Due to this high ground-state energy of $^{25}$O it is likely that $^{26}$O is bound with respect to one-neutron emission and unbound with respect to two-neutron emission. $^{26}$O is thus also an excellent candidate for di-neutron emission. Furthermore calculations by Grigorenko {\it et al.} predict that the emission of a pair of correlated neutrons might be hindered so that for very low decay energies lifetimes on the order of pico- to nanoseconds could be possible \cite{2011Gri01}.

\begin{figure}[t]
\includegraphics[width=0.35\textwidth]{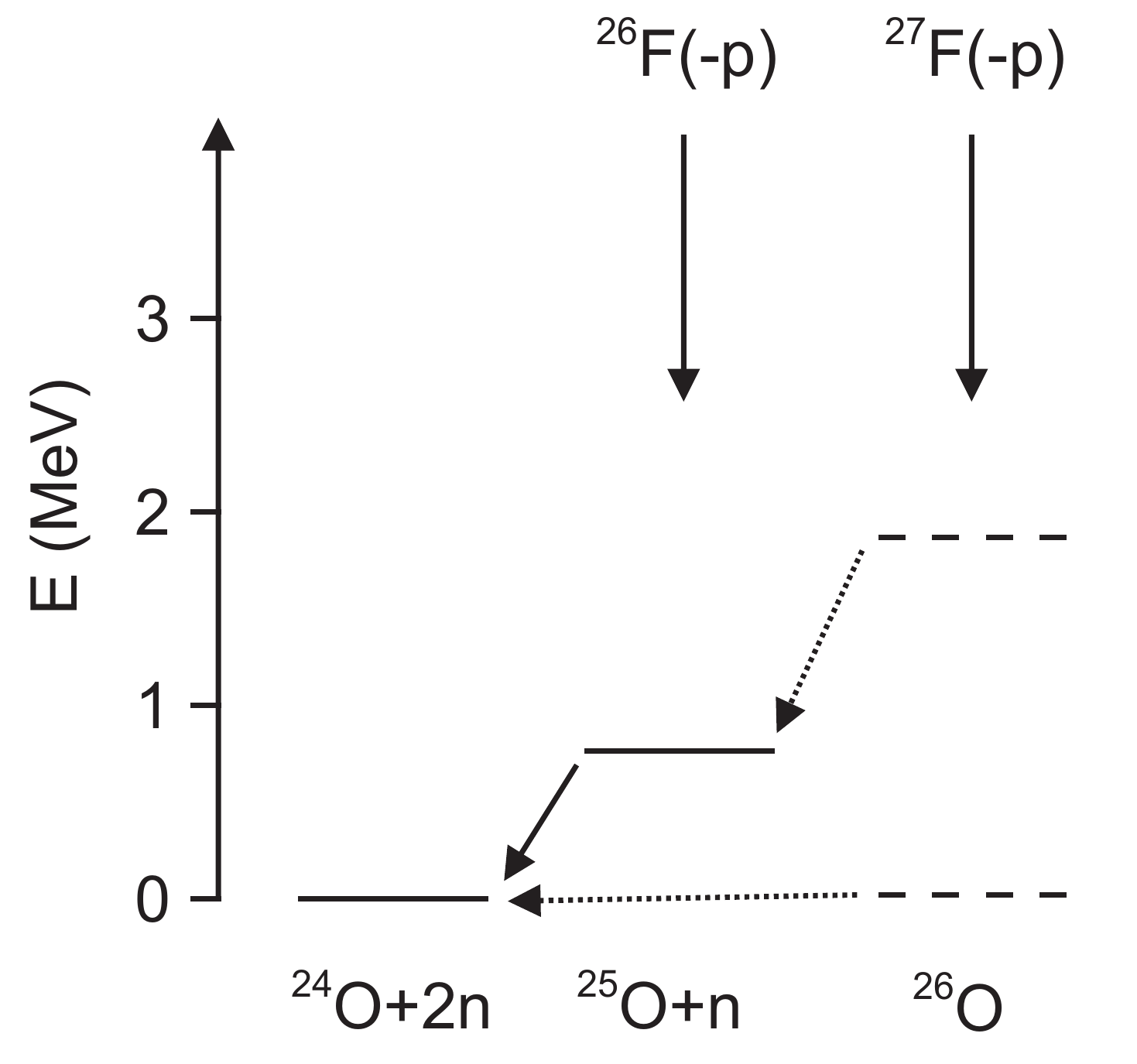}
\caption{\label{fig:levels} Schematic decay level scheme of $^{26}$O. Proton-knockout reactions from $^{26}$F populate the ground state of $^{25}$O which was measured to decay to the ground state of $^{24}$O (solid arrow and lines) \cite{2008Hof01}. The knockout reaction from $^{27}$F used in the present work populates states in $^{26}$O. The dashed lines show the predicted levels calculated by the continuum shell model \cite{2006Vol01}. Possible decay channels from $^{26}$O are shown by the dotted arrows.}
\end{figure}

We searched for unbound states in $^{26}$O using one-proton knockout reactions from $^{27}$F and by measuring neutrons in coincidence with $^{24}$O fragments. Figure \ref{fig:levels} shows a schematic level scheme of the possible decay paths for predicted states of $^{26}$O. In this letter we present the first evidence for the observation of the unbound ground state of $^{26}$O.

\begin{figure}[t]
\includegraphics[width=0.4\textwidth]{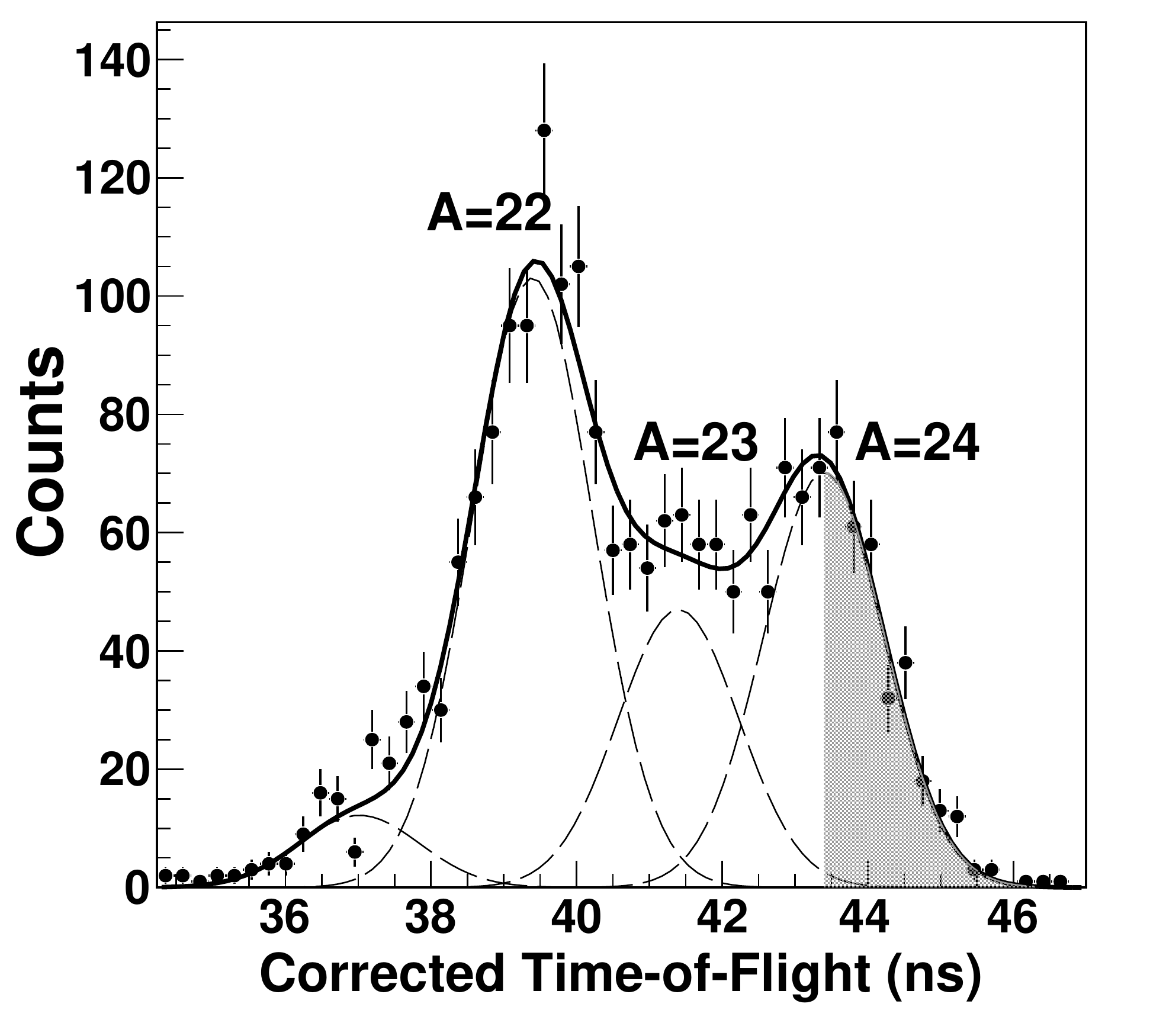}
\caption{\label{fig:id} Particle identification spectrum. The three oxygen isotopes $^{22}$O, $^{23}$O, and $^{24}$O were identified as indicated in the figure. The grey-shaded area corresponds to the events included in the present analysis for the decay of $^{25}$O and $^{26}$O.}
\end{figure}

The experiment was performed at the Coupled Cyclotron Facility of the National Superconducting Cyclotron Laboratory at Michigan State University. A primary beam of 140~MeV/u $^{48}$Ca$^{20+}$ bombarded a 1316~mg/cm$^2$ $^9$Be production target. The desired $^{27}$F secondary beam with an energy of 82~MeV/u was separated from the other reaction products, and primary beam, using the A1900 fragment separator \cite{A1900} with a 1050~mg/cm$^2$ Al wedge placed at the intermediate focal plane.  The $^{27}$F component of the secondary beam was 7\% with the main contaminant being $^{30}$Na. The $^{27}$F fragments were identified event-by-event through time-of-flight, and the average intensity was 14 per second  with a momentum spread of 2.5\%. The $^{27}$F beam then impinged on a 705~mg/cm$^{2}$ Be target producing the isotope of interest, $^{26}$O, through one-proton removal reactions.  Measurement of the $^{26}$O $\rightarrow$ $^{24}$O + 2n decay required detection of both the neutrons and a charged particle.  Beyond the reaction target, the large-gap 4~Tm superconducting dipole (Sweeper) magnet \cite{SWEEPER} was used to bend the charged particles 43$^{\circ}$ after which they passed through a set of position and energy sensitive detectors.  The $^{24}$O fragments were identified by time-of-flight and energy loss in the charged particle detectors.  The Modular Neutron Array (MoNA) \cite{MONA} was placed 6.05~m from the reaction target and measured the angle and energy (from time-of-flight) of the forward-focused beam-velocity neutrons.  The charged particle detectors and MoNA provide the relativistic four-momentum vectors for the $^{24}$O nuclei and neutrons that were used to calculate the decay energy of the two-body (fragment + n) or three-body (fragment + n + n) systems. The thicker target, as compared to previous experiments, increases the FWHM of the decay energy resolution to 200 and 800~keV for decay energies of 100 and 800~keV, respectively. Further details about the experimental set-up, parameters and analysis procedures can be found in references \cite{2008Hof01,SCHILLER07,FRANK07,FRANK08,2009Hof01}.

\begin{figure}[t]
\includegraphics[width=0.4\textwidth]{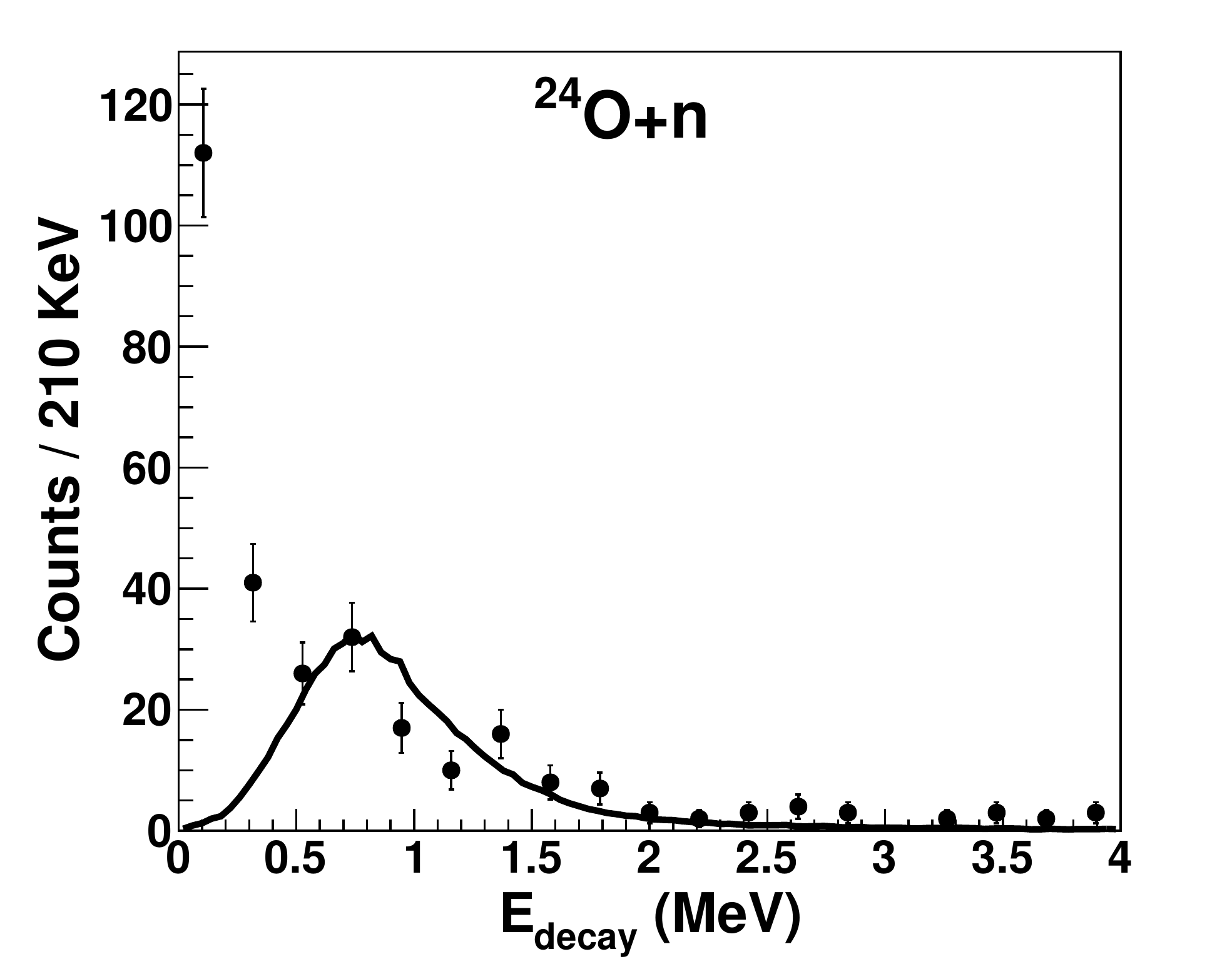}
\caption{\label{fig:calem} Decay energy spectrum of $^{25}$O. The data points are from the present one-proton knockout reaction from $^{27}$F, while the solid line corresponds to the fit to the data of the one-proton knockout reaction from $^{26}$F from reference \cite{2008Hof01}.}
\end{figure}

Figure \ref{fig:id} shows the particle identification spectrum for the oxygen isotopes. The events included in the further analysis are shown by the grey-shaded area. With this stringent cut the contamination of $^{23}$O events was limited to about 1\%.

\begin{figure*}[t]
\includegraphics[width=0.8\textwidth]{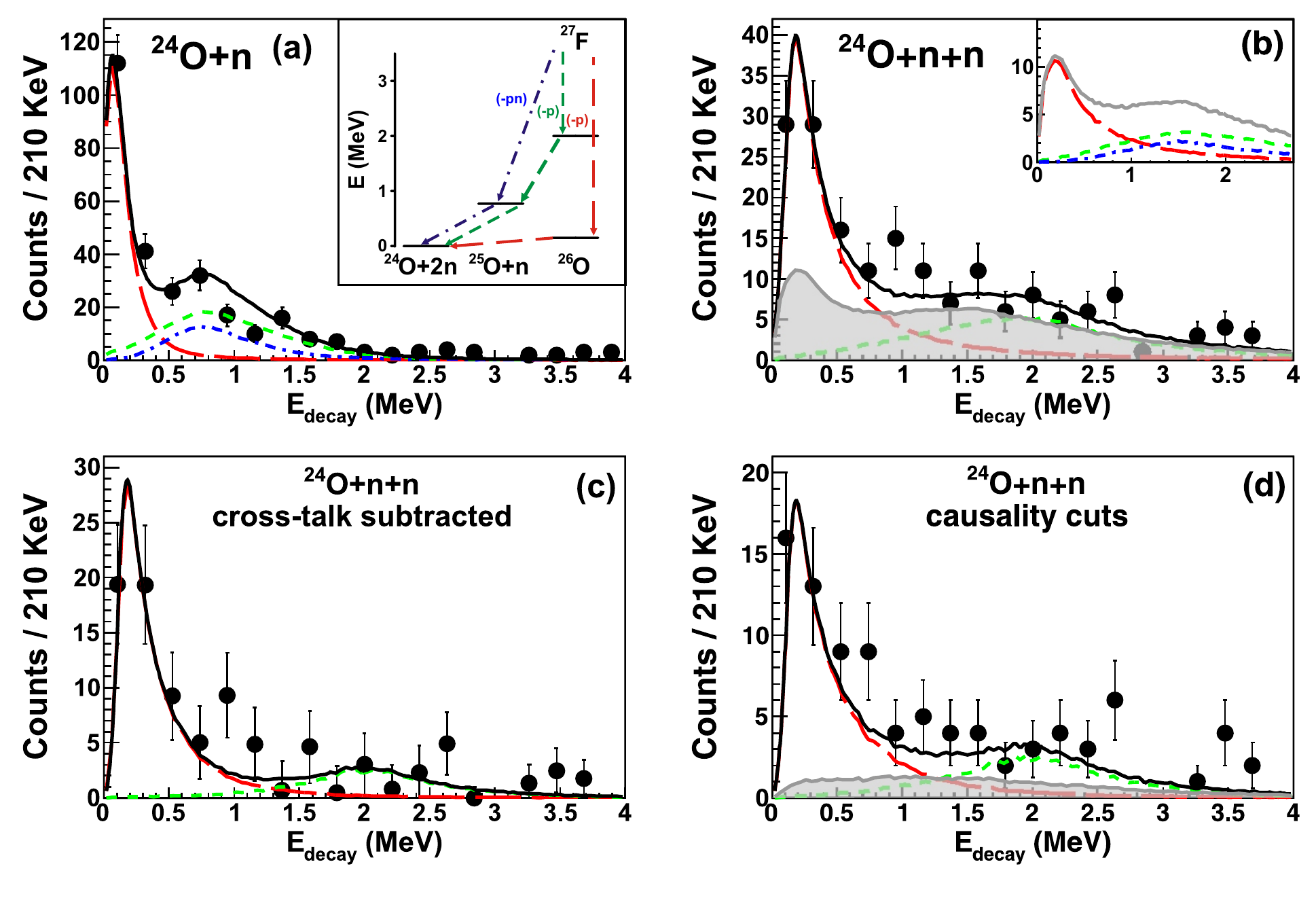}
\caption{\label{fig:26O} (Color Online) (a) Decay energy spectrum of $^{25}$O (two-body, fragment + n system). The data are the same as in Figure \ref{fig:calem}. The inset indicates the different decay paths. (b,c,d) decay energy spectra of $^{26}$O (three-body, fragment + n + n system). The grey area in (b) represents the simulated cross-talk contribution. The individual contributions to the cross-talk from the different simulated states are shown in the inset. Panel (c) shows the cross-talk subtracted $^{26}$O spectrum. Panel (d) shows the $^{26}$O spectrum with causality cuts applied to the data as well as the simulation. The lines are explained in the text.}
\end{figure*}

The decay energy spectrum for the two-body (fragment + n) system corresponding to $^{25}$O is shown in Figure \ref{fig:calem}. $^{25}$O previously had been measured and exhibited a resonance with a decay energy of 770$^{+20}_{-10}$~keV \cite{2008Hof01}. The solid line is representative of this resonance. It is immediately obvious that the data from the present one-proton knockout reaction from $^{27}$F (data points) has a completely different shape with an additional strong peak at low decay energies. Because these low-decay energy neutrons cannot come from $^{25}$O they have to originate from $^{26}$O. This initial evidence for a low energy resonance in $^{26}$O is substantiated by the presence of a low-energy peak in the reconstructed three-body (fragment + n + n) system corresponding to $^{26}$O as shown in Figure \ref{fig:26O} (b). It should be mentioned that in the present experiment as well as in the experiment of reference \cite{2009Hof01} data were also recorded for neutrons in coincidence with $^{23}$O + n, and the measured decay energy spectra from the two experiments are in agreement.

The three-body decay energy spectrum shown in Figure \ref{fig:26O} (b) was reconstructed from a $^{24}$O fragment and two neutron interactions in MoNA. Thus, the spectrum also contains cross-talk events from a single neutron scattering twice. The broad distribution of counts up to approximately 3~MeV in the three-body decay energy spectrum could be due to the ground state of $^{25}$O where the neutron scattered twice and/or the sequential decay from excited states in $^{26}$O through the $^{25}$O ground state. In order to fit the data, we performed Monte Carlo simulations which included the geometrical acceptances, energy, positions, and timing resolution, tracking of the charged particles through the Sweeper magnet, and the reaction and decay mechanisms.  The interaction of the neutrons with MoNA was described using the Geant4 simulation package~\cite{GEANT} with the addition of the MENATE-R physics class \cite{MENATE}. Thus, multiple interaction of a single neutron were fully included in the simulations.

A low-energy resonance was simulated with a Breit-Wigner line shape and the energy from the $^{26}$O decay into $^{24}$O and two neutrons was partitioned between the three outgoing particles according to the phase-space model of references \cite{PHASESPACE,ROOT}. The data are not sensitive to the detailed parameters of the various possible contributions to the high energy continuum. As shown in Figure \ref{fig:levels}, the continuum shell model predicts only one excited state for $^{26}$O at approximately 2~MeV with the next excited state calculated at about 6~MeV \cite{2006Vol01}. Thus, a resonance in $^{26}$O at a fixed energy of 2~MeV with a width of 200~keV, which was allowed to decay sequentially via the known unbound ground state of $^{25}$O, was included in the simulation. The resonance parameters for $^{25}$O ($E_{\text{decay}} = 770$~keV, $\Gamma_{\text{decay}} = 172$~keV, $L = 2$) were taken from reference \cite{2008Hof01}. A $\chi^2$ fit to the two-body and three-body system was performed where the resonance energy and width of the low-energy resonance and the relative strengths of the three contributions (low-energy and 2~MeV state in $^{26}$O and the ground state of $^{25}$O) were free parameters.

Figure \ref{fig:26O} shows the resulting simulated spectra for the best fit parameters (solid line). The two-body and three-body systems are shown on panels (a) and (b), respectively. The low-energy resonance in $^{26}$O at $E_{\text{decay}} = 150$~keV and $\Gamma_{\text{decay}} = 5$~keV is shown by the long-dashed red line and the 2~MeV resonance by the short-dashed green line. The dot-dashed blue line shows the contribution from the direct population of the $^{25}$O ground state. These decay paths are also indicated in the inset of Figure \ref{fig:26O}(a). In the simulation it is possible to distinguish real two-neutron detection from cross-talk events where a single neutron interacted twice in MoNA. The grey-shaded area in Figure \ref{fig:26O} (b) shows the contributions of the simulated cross-talk events to the total spectrum. In the inset the individual contributions to the cross-talk from the low- and high-energy $^{26}$O decays as well as the contribution from the $^{25}$O are shown. In order to demonstrate the positive signal of real 2n events we applied two different methods. First we subtracted the simulated cross-talk events from the data, and the results are shown in Figure \ref{fig:26O}(c). While the high-energy events are essentially reduced to background level, the low-energy peak is clearly still present. In a second method we applied causality cuts to the data. Following the description of references \cite{2006Nak01,2011Hof01} we required a spatial (25~cm) and velocity (7~cm/ns) separation of two interactions in the data as well as in the simulation. At the expense of a reduction of $\sim$50\% in efficiency, the cross-talk contribution was reduced by a factor of three. The results are shown in figure \ref{fig:26O}(d) which again clearly shows the presence of the low energy peak of real two-neutron events.

The best fit to the data included a resonance for the $^{26}$O ground state of 150$^{+50}_{-150}$~keV. This value agrees with the recent calculations of a low-energy unbound resonance in $^{26}$O \cite{2006Vol01,2010Ots01}. The fit was insensitive to the width of the resonance. The cross section for populating the $^{26}$O ground state was 1.8$\pm$1.0~mb while the cross section for populating $^{25}$O was 4.2$\pm$2~mb. These values are consistent with 1p and 1p1n removal cross sections from $^{24}$F, $^{25}$F and $^{26}$F. While the measured values for $^{24}$F agreed with the calculated removal cross sections the cross sections decreased with increasing neutron number \cite{2003Tho01}.

\begin{figure}[t]
\includegraphics[width=0.4\textwidth]{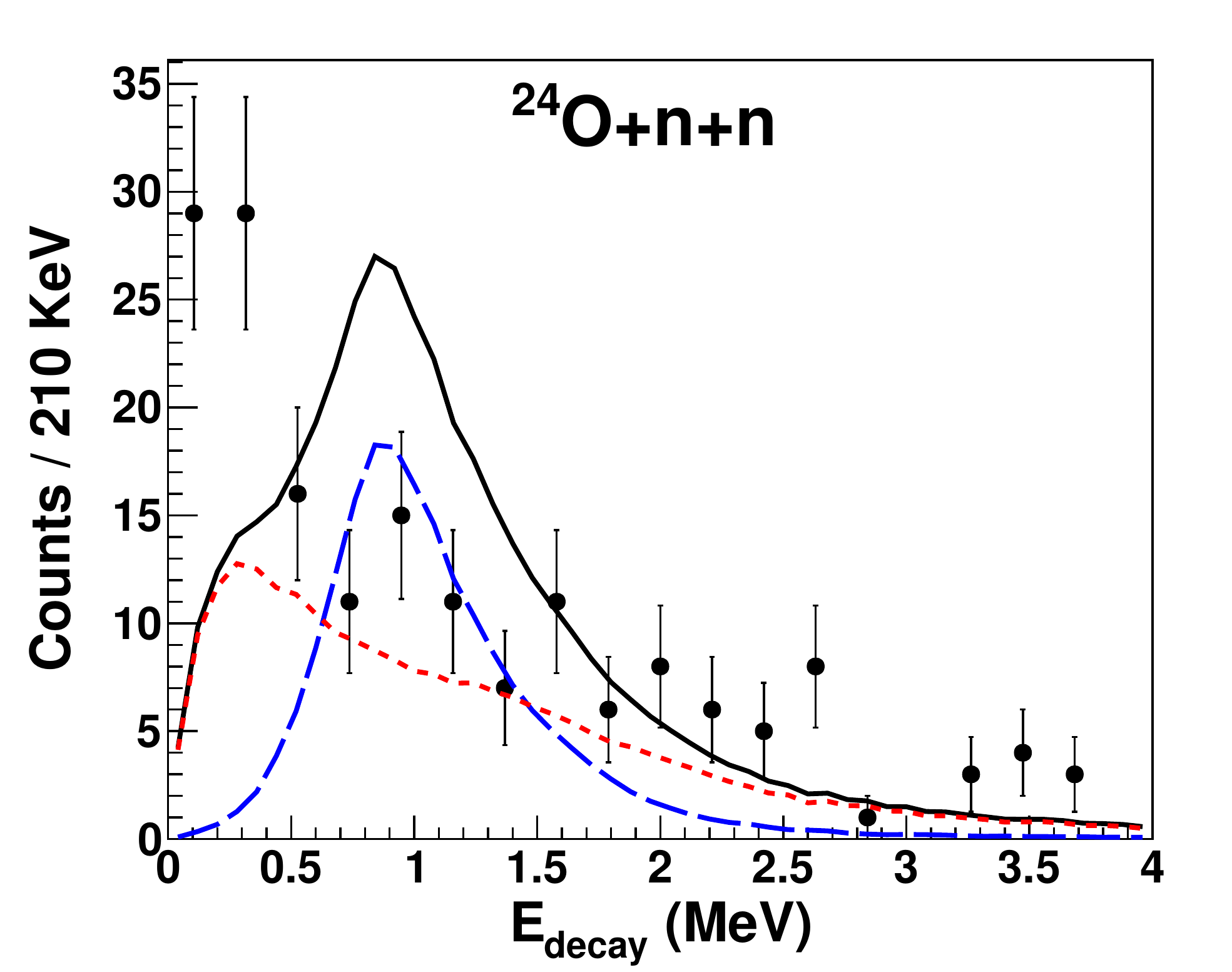}
\caption{\label{fig:seq} (Color online) Decay energy spectrum from two neutron interactions in coincidence with $^{24}$O (same as Figure \ref{fig:26O}(b)). The solid line corresponds to the results of a simulation assuming a hypothetical 870~keV state in $^{26}$O decaying sequentially via the ground state of $^{25}$O. The dashed and dot-dashed lines are the contributions due to real two-neutron coincidences and double-hits from one neutron, respectively.}
\end{figure}

One might speculate that the low-energy neutron originates from a state located around 900~keV and then decays with a $\sim$100~keV neutron sequentially via the ground state of $^{25}$O. The observed low energy peak would then be solely due to the low-energy neutron scattering twice in MoNA. We simulated such a decay and the results are shown in Figure \ref{fig:seq}. In addition to the total fit (solid line), the figure also shows the individual contributions from real two-neutron coincidences (long-dashed blue line) and from events where either the low- or the high-energy neutron scatters twice in MoNA (short-dashed red line). Clearly, this sequential decay scenario is not supported by the data.

In conclusion, we present evidence for the observation of the $^{26}$O ground state which is unbound by less than 200~keV and which decays by emitting two low-energy neutrons. A future experiment with higher statistics is necessary to study the detail of the decay mechanism, i.e. explore the possibility of a di-neutron decay. The upper limit for the total decay energy leaves open the exciting prospect that a di-neutron decay of $^{26}$O might have a long half-life ($\gtrsim$ picoseconds) as first suggested by Grigorenko {\it et al.} \cite{2011Gri01}.

The authors gratefully acknowledge the support of the NSCL operations staff for providing a high quality beam and D. Sayre for taking night shifts during the experiment. This work was supported by the National Science Foundation under Grants
PHY-02-44953, 
PHY-08-55456, 
PHY-09-69058, and 
PHY-06-06007. 





\end{document}